# Nonuniqueness of canonical ensemble theory arising from microcanonical basis

Sumiyoshi Abe[1] and A. K. Rajagopal[2]

[1]*College of Science and Technology, Nihon University,*

*Funabashi, Chiba 274-8501, Japan*

[2]*Naval Research Laboratory, Washington, D.C. 20375-5320, USA*

Given physical systems, counting rule for their statistical mechanical descriptions need not be unique, in general. It is shown that this nonuniqueness leads to the existence of various canonical ensemble theories, which equally arise from the definite microcanonical basis. Thus, the Gibbs theorem for canonical ensemble theory is not universal, and maximum entropy principle is to be appropriately modified for each physical context.

It is generally believed that canonical ensemble theory arising from the microcanonical basis is unique. This conviction seems to have its origins in the demonstration given by Gibbs [1]. In this Letter, we show that it is actually not true. A crucial point regarding this remarkable fact is that, given a system, rule for counting its physical configurations is not unique and is context-dependent, in general.

A counting rule is an algorithm, which establishes a connection between entropy, a macroscopic quantity, and the number of microscopic configurations of the system under consideration. The most celebrated rule is that of Boltzmann, who employed the logarithmic function for counting the microscopic configurations. A wide class of physical systems is covered by this logarithmic counting rule. However, strictly speaking, there is no *a priori* reason to assume such a special counting rule to be completely universal. In fact, there may yet exist a class of systems which seems to prefer alternate kinds of counting rules. Specifically, nonextensive systems, such as systems with long-range interactions, long-range memories or (multi)fractal structure, belong to this class. In what follows, we first summarize the derivation of ordinary canonical ensemble theory from microcanonical basis with the principle of equal *a priori* probability based on the logarithmic counting rule. Then, we present a new counting rule, which leads to a non-Gibbsian canonical ensemble, showing the nonuniqueness of canonical ensemble theory.

Consider a system $s$ and take $N$ replicas $s_1, s_2, \mathrm{L}, s_N$ of it. $\mathrm{S} = \{s_\alpha\}_{\alpha=1,2,\mathrm{L},N}$ is referred to as a supersystem. Let $A_\alpha$ be a physical quantity associated with $s_\alpha$ and is assumed to be bounded from below. It can be thought of as the system energy, for example. Its value is denoted by $a(m_\alpha)$, where $m_\alpha$ labels the allowed configurations of $s_\alpha$. A quantity of interest is given by its average over the supersystem: $(1/N)\sum_\alpha A_\alpha$. Microcanonical ensemble theory states that the probabilities of finding $\mathrm{S}$ in the configurations where the values of the macroscopic quantity lies around a certain value $\bar{a}$, i.e.

$$\left| \frac{1}{N} \sum_{\alpha=1}^{N} a(m_\alpha) - \bar{a} \right| < \varepsilon \qquad (1)$$

are all equal. Here, $\varepsilon$ goes to zero with $N \to \infty$. The ordinary central limit theorem and law of large numbers indicate that $\varepsilon \sim O(1/\sqrt{N})$. However, if the Lévy-type power-law canonical distribution is in mind, relevant mathematics is the Lévy-Gnedenko generalized central limit theorem, and accordingly the assumption on the order of $N$-dependence of $\varepsilon$ should appropriately be modified.

In Ref. [2], a way of deriving ordinary canonical ensemble theory is presented based on the logarithmic counting rule. The probability $p(m)$ of finding the system $s \equiv s_1$ in its $m$th configuration is given by the ratio of two numbers of equally-probable configurations $\{m_1, m_2, \mathrm{L}, m_N\}$, i.e. $p(m) = W(m)/W$. Here, $W$ is the total number of configurations satisfying eq. (1), whereas $W(m)$ is determined by the following two conditions: a) $s_1$ is found in $m_1 = m$, and b) $\bar{a}$ in the condition in eq. (1) is the arithmetic mean over the configurations of $S$. To calculate it explicitly, we rewrite eq. (1) as follows:

$$\left| \frac{1}{N}[a(m) - \bar{a}] + \frac{1}{N} \sum_{\alpha=2}^{N} a(m_\alpha) - \frac{N-1}{N} \bar{a} \right| < \varepsilon. \qquad (2)$$

The number of configurations $Y_N$ satisfying

$$\left| \frac{1}{N} \sum_{\alpha=2}^{N} a(m_\alpha) - \frac{N-1}{N} \bar{a} \right| < \varepsilon \qquad (3)$$

is counted in the large-$N$ limit as follows:

$$\ln\left[Y_N\left(\frac{N-1}{N}\bar{a}\right)\right]^{\frac{1}{N}} \cong \ln[Y_N(\bar{a})]^{\frac{1}{N}} \to S(\bar{a}), \qquad (4)$$

where $S(\bar{a})$ is some function of $\bar{a}$ and will be identified with the entropy. This logarithmic counting rule imposes the *additional* extensivity assumption on the entropy. From eqs. (2)-(4), we have

$$\ln W(m) = \ln Y_N\left(\frac{N-1}{N}\bar{a} - \frac{1}{N}[a(m)-\bar{a}]\right) \cong \ln Y_N(\bar{a}) - [a(m)-\bar{a}]\frac{\partial S}{\partial \bar{a}}. \qquad (5)$$

Therefore, defining $\beta$ as

$$\beta = \frac{\partial S(\bar{a})}{\partial \bar{a}} \qquad (6)$$

and putting

$$\tilde{Z}(\beta) = \lim_{N \to \infty} \frac{W}{Y_N(\bar{a})}, \qquad (7)$$

we obtain the canonical distribution

$$p(m) = \frac{1}{\tilde{Z}(\beta)}\exp\{-\beta[a(m)-\bar{a}]\}, \qquad (8)$$

where $\tilde{Z}(\beta) = Z(\beta)\exp(\beta\bar{a})$ with the ordinary canonical partition function $Z(\beta) = \sum_m \exp[-\beta a(m)]$.

In maximum entropy principle [3], the distribution in eq. (8) is derived from the Boltzmann-Shannon entropy

$$S = -\sum_m p(m)\ln p(m) \qquad (9)$$

under the constraints on the normalization of $p(m)$ and the expectation value

$$\sum_m p(m)a(m) = \bar{a}. \qquad (10)$$

Then, $\beta$ is the Lagrange multiplier associated with this constraint. Note that equation (10) is equivalent to the condition

$$\left.\frac{\partial \tilde{Z}(\beta)}{\partial \beta}\right|_{\bar{a}} = 0. \qquad (11)$$

On the other hand, the entropy in eq. (9) is calculated to be

$$S = \ln \tilde{Z}(\beta). \qquad (12)$$

Therefore, equation (11) is actually the maximum entropy condition. Consistency of eq. (12) with eq. (4) identifies

$$\tilde{Z}(\beta) = \lim_{N \to \infty} [Y_N(\bar{a})]^{\frac{1}{N}}. \tag{13}$$

Also, from eq. (7), $W$ is found to be

$$W = \lim_{N \to \infty} [Y_N(\bar{a})]^{\frac{1}{N}+1}. \tag{14}$$

In the above derivation, the logarithmic counting in eq. (4) plays an essential role. Now, let us examine another kind of counting rule. Specifically, we examine to replace eq. (4) by

$$\ln_q \left[ Y_N \left( \frac{N-1}{N} \bar{a} \right) \right]^{\frac{1}{N}} \cong \ln_q [Y_N(\bar{a})]^{\frac{1}{N}} \to S_q(\bar{a}). \tag{15}$$

Here, $\ln_q x$ is the "$q$-logarithm" defined by

$$\ln_q x = \frac{1}{1-q}(x^{1-q} - 1), \tag{16}$$

which is the inverse function of the "$q$-exponential"

$$e_q(x) = \begin{cases} [1+(1-q)x]^{\frac{1}{1-q}} & (1+(1-q)x > 0) \\ 0 & (\text{otherwise}), \end{cases} \tag{17}$$

where $q$ is a positive parameter. In the limit $q \to 1$, $\ln_q x$ and $e_q(x)$ converge to the ordinary $\ln x$ and $e^x$, respectively. As the ordinary logarithm, $\ln_q x$ is also a monotonically increasing function of $x$. $S_q(\bar{a})$ in eq. (15) is a certain quantity dependent on $\bar{a}$ and will be *shown* to be the Tsallis nonextensive entropy [4]. Instead of eq. (5), now our algorithm is given as follows:

$$\ln_q W(m) = \ln_q Y_N\left(\frac{N-1}{N}\bar{a} - \frac{1}{N}[a(m)-\bar{a}]\right)$$

$$\cong \ln_q Y_N(\bar{a}) - \frac{1}{N}[a(m)-\bar{a}]\frac{\partial}{\partial \bar{a}}\ln_q Y_N(\bar{a}), \qquad (18)$$

provided that $N$ may be large but finite. Defining

$$\frac{\partial S_q(\bar{a})}{\partial \bar{a}} = \beta \qquad (19)$$

and using eq. (15), we have

$$\beta = \frac{1}{N}[Y_N(\bar{a})]^{\frac{1-q}{N}-1}\frac{\partial Y_N(\bar{a})}{\partial \bar{a}}. \qquad (20)$$

Therefore, we find

$$\frac{\partial}{\partial \bar{a}}\ln_q Y_N(\bar{a}) = \beta N [Y_N(\bar{a})]^{(1-q)\left(1-\frac{1}{N}\right)}. \qquad (21)$$

Substituting eq. (21) into eq. (18), we have

$$\ln_q W(m) - \ln_q Y_N(\bar{a}) = -\beta [Y_N(\bar{a})]^{(1-q)\left(1-\frac{1}{N}\right)}[a(m) - \bar{a}]. \tag{22}$$

Taking advantage of the relation

$$\ln_q \left(\frac{x}{y}\right) = y^{q-1}\left(\ln_q x - \ln_q y\right), \tag{23}$$

we obtain

$$\frac{W(m)}{Y_N(\bar{a})} \cong e_q\left(-\beta^*[a(m) - \bar{a}]\right), \tag{24}$$

where

$$\beta^* \equiv \frac{\beta}{c_q} \tag{25}$$

with

$$c_q \equiv [Y_N(\bar{a})]^{\frac{1-q}{N}} = 1 + (1-q)S_q(\bar{a}). \tag{26}$$

From eq. (24) and the total number of configurations $W = \sum_m W(m)$, we construct

$$\tilde{Z}_q(\beta) \equiv \frac{W}{Y_N(\bar{a})} \cong \sum_m e_q\left(-\beta^*[a(m) - \bar{a}]\right). \tag{27}$$

Consequently, we find the probability $p_q(m) = W(m)/W$ to be

$$p_q(m) = \frac{1}{\tilde{Z}_q(\beta)} e_q\left(-\beta^*[a(m) - \bar{a}]\right). \tag{28}$$

The property corresponding to eq. (11) here translates to

$$\left.\frac{\partial \tilde{Z}_q(\beta)}{\partial \beta}\right|_{\bar{a}} = 0, \tag{29}$$

leading to the result

$$\bar{a} = <A>_q \equiv \sum_m P_q(m) a(m), \tag{30}$$

where $P_q(m)$ is the escort distribution [5]

$$P_q(m) = \frac{[p_q(m)]^q}{\sum_m [p_q(m)]^q}. \tag{31}$$

Equation (30) shows that the arithmetic mean is identical with the normalized $q$-expectation value introduced in Ref. [6].

Now, we show that consistency of the whole discussion is achieved if the generalized entropy is given by

$$S_q = \ln_q \tilde{Z}_q(\beta), \tag{32}$$

which should be compared with eq. (12). Accordingly, equation (29) becomes the maximum generalized entropy condition. Consistency between eqs. (32) and (15) leads to the identification

$$\tilde{Z}_q(\beta) \cong [Y_N(\bar{a})]^{\frac{1}{N}}, \tag{33}$$

as in eq. (13). Furthermore, from eq. (27), $W$ is seen to be

$$W \cong [Y_N(\bar{a})]^{\frac{1}{N}+1}, \tag{34}$$

which is formally similar to eq. (14).

Now, note that the identical relation

$$\sum_m [p_q(m)]^q = [\tilde{Z}_q(\beta)]^{1-q} \tag{35}$$

holds for the distribution in eq. (28) with eq. (30). Combining this with eq. (33), we have

$$[Y_N(\bar{a})]^{\frac{1-q}{N}} = \sum_m [p_q(m)]^q. \tag{36}$$

Therefore, $c_q$ in eq. (26) is found to be

$$c_q = \sum_m [p_q(m)]^q. \tag{37}$$

Finally, using eqs. (33), (36) and (37), we ascertain that both eqs. (26) and (32) consistently lead to

$$S_q = \frac{1}{1-q} \left\{ \sum_m [p_q(m)]^q - 1 \right\}, \qquad (38)$$

which is precisely the Tsallis entropy [4].

In the limit $q \to 1$, eqs. (28), (30) and (38) converge to eqs. (8), (10) and (9), respectively.

We end our discussion with emphasizing the following points. $p_q(m)$ in eq. (28) is the generalized canonical distribution in nonextensive statistical mechanics [6], which is derived via the maximum entropy principle based on the Tsallis entropy in eq. (38) subject to the constraints on the normalization of $p(m)$ and the normalized $q$-expectation value of $A$ given in eq. (30) and is known to lead to the consistent thermodynamic formalism. $\beta$ in eq. (19) is the Lagrange multiplier associated with eq. (30). Another important point is that if the limit $N \to \infty$ is strictly taken, for example, in eq. (27), then the difference between $\tilde{Z}(\beta)$ in eq. (7) and $\tilde{Z}_q(\beta)$ in eq. (27) may disappear. In other words, $q$ seems to go to unity in the limit $N \to \infty$. This suggests that nonuniqueness of counting rules may be intimately related to the fact that physically $N$ is large but actually finite. Such an observation is consistent with the discussion developed in Refs. [7,8]. This leads to a fundamental question regarding a connection between the concepts of extensivity and thermodynamic limit. These considerations, in turn, indicate that nonextensive statistical mechanics would also be important for understanding thermodynamic properties of small systems.

In conclusion, we have found that canonical ensemble theory arising from the conventional microcanonical basis with the principle of equal *a priori* probability is not unique. We have shown the nonuniqueness by employing a non-logarithmic counting rule, the $q$-logarithmic counting, as an example. We have found that, in this case, the resulting canonical ensemble theory is Tsallis' nonextensive formalism. We stress that the concept of the arithmetic mean is the same throughout the discussion: the nonuniqueness is brought about through counting rule and concomitant definition of the statistical expectation value. We wish to mention that the same result is obtained by using the method of steepest descents [9] and the Lévy-Gnedenko generalized central limit theorem [10].

The authors would like to thank Professor Roger Balian for drawing their attention to Ref. [2], where discussions about uniqueness of the Gibbs canonical ensemble theory are developed. The present work arose from a careful reexamination of Ref. [2]. S. A. was supported in part by the GAKUJUTSU-SHO Program of College of Science and Technology, Nihon University. A. K. R. acknowledges the support of the U.S. Office of Naval Research.